%% file: fwgan.tex
\documentclass{article}
\usepackage{iftex}
\ifTUTeX
  \usepackage{fontspec}
\else
  \usepackage[T1]{fontenc}
  \usepackage[utf8]{inputenc} 
  \DeclareUnicodeCharacter{200B}{{\hskip 0pt}}
\fi
\usepackage{spconf,amsmath,graphicx}
\usepackage{caption}
\usepackage{subcaption}
\usepackage{siunitx}
\usepackage{comment}
\usepackage{balance}

\usepackage{bm}
\usepackage{amsfonts}
\usepackage{graphicx}
\usepackage{amsmath}
\usepackage{multirow}
\usepackage{textcomp}
\usepackage{tabularx,booktabs}
\usepackage{makecell}
\usepackage{footnote}
\usepackage[hyphens,spaces,obeyspaces]{url}
\usepackage[hidelinks]{hyperref}
\title{Framewise WaveGAN: High Speed Adversarial Vocoder in Time Domain with Very Low Computational Complexity}
\name{Ahmed Mustafa$^{\star}$ \qquad Jean-Marc Valin$^{\star}$  \qquad Jan B{\"u}the$^{\star}$  \qquad Paris Smaragdis$^{\star \dagger}$  \qquad Mike Goodwin$^{\star}$}

\address{$^{\star}$Amazon Web Services, Palo Alto, USA\\
$^{\dagger}$University of Illinois at Urbana-Champaign\\
\texttt{\{ahdmust, jmvalin, jbuethe, parsmara, mmg\}@amazon.com}}

\begin{document}
\topmargin=0mm
\ninept
\maketitle
\begin{abstract}
GAN vocoders are currently one of the state-of-the-art methods for building high-quality neural waveform generative models. However, most of their architectures require dozens of billion floating-point operations per second (GFLOPS) to generate speech waveforms in samplewise manner. This makes GAN vocoders still challenging to run on normal CPUs without accelerators or parallel computers. In this work, we propose a new architecture for GAN vocoders that mainly depends on recurrent and fully-connected networks to directly generate the time domain signal in framewise manner. This results in considerable reduction of the computational cost and enables very fast generation on both GPUs and low-complexity CPUs. Experimental results show that our Framewise WaveGAN vocoder achieves significantly higher quality than auto-regressive maximum-likelihood vocoders such as LPCNet at a very low complexity of \SI{1.2}{GFLOPS}. This makes GAN vocoders more practical on edge and low-power devices.
\end{abstract}
%

\section{Introduction}
\label{SEC_intro}

\input{sections/introduction}


\section{Framewise WaveGAN}
\label{fwgan}

\input{sections/fwgan}

\section{Training Procedure}
\label{train}
\input{sections/training}

\section{Experiments and Results}

\input{sections/experiments}

\section{Conclusion}
\label{SEC_conclusion}
This work introduced Framewise WaveGAN, an adversarial vocoder for wideband speech synthesis with very low complexity, close to \SI{1}{GFLOPS}, and a low delay of \SI{25}{ms}. The model architecture comprises recurrent and fully-connected layers that capture long- and short-term dependencies of the speech signal, with ability to generate the waveform frame-by-frame faster than real-time. Quality evaluations show that Framewise WaveGAN significantly outperforms state-of-the-art auto-regressive vocoders such as LPCNet. Potential future work includes better adversarial training methods to boost this kind of framewise vocoder in achieving high-fidelity speech generation at low delay and complexity. 

\section{Acknowledgments}
\label{SEC_ack}
We would like to thank Umut Isik, Timothy B. Terriberry, Karim Helwani and Erfan Soltanmohammadi for their kind support and fruitful tips.

\balance
\bibliographystyle{IEEEbib}
\bibliography{fwgan}

\end{document}

%% file: sections/introduction.tex
The task of neural vocoding has seen rapid progress in recent years. This came with lots of applications that depend on neural vocoders for rendering high quality speech; such as TTS~\cite{tan2021survey}, speech enhancement~\cite{9632770} and neural speech coding~\cite{mustafa2021streamwise, zeghidour2021soundstream, pia2022nesc}. 
Since the advent of Wavenet~\cite{wavenet}, deep generative models have become the standard way for building high quality neural vocoders, with clear outperformance over classical vocoding methods. By modelling raw waveform data distributions in the time domain, deep generative models can capture intricate representational details of the speech signal, which are not easy to track by fully-parametric methods. This results in generating high quality speech signals; but it also incurs huge amount of computations that scales rapidly with the target sampling rate. In case of WaveNet, this high computational cost becomes more challenging due to the auto-regressive (AR) sampling that makes the overall generation very slow. Later models such as WaveRNN~\cite{wavernn} tried to address one part of this problem by using a sparse RNN that runs auto-regressively to learn the data distribution. LPCNet~\cite{lpcnet} further improved WaveRNN by using linear prediction to build lighter architectures while maintaining or even enhancing the signal quality; and currently it is one of the lowest complexity generative models available for raw audio. Further AR models~\cite{vipperla2020bunched,tian2020featherwave} follow the same approach but generate more than one sample at a time to speed up the inference process. 

Parallel generative models propose a different way for dealing with the high computational cost. These models avoid auto-regressive generation to parallelize computations via GPUs or CPU accelerators. There are many types of parallel vocoders which are similar in their architecture but differ in the way they are trained. GAN vocoders~\cite{kong2020hifi, mustafa2021stylemelgan, 9746675} are more competitive in parallel generation due to their light-weight models, as compared to maximum-likelihood (ML) vocoders like WaveGlow~\cite{prenger2019waveglow}, and their one-shot generation, as opposed to diffusion vocoders~\cite{kong2020diffwave, koizumi2022specgrad} that usually require iterative denoising steps to achieve high quality signal generation. However, despite their high quality and very fast signal generation, GAN vocoders still require multiples of Giga Floating-Point Operations per Second (GFLOPS), especially when using WaveNet-based architectures. This amount of computation is challenging for low complexity devices with limited power and parallelization capabilities. 

Our work aims to reduce the large computational amount of current GAN vocoders. Specifically, we report the following contributions:
\begin{itemize}
\item We introduce Framewise WaveGAN, a neural vocoder generating wideband speech signals in the time domain frame-by-frame instead of sample-by-sample.
\item We show how to train the model reliably and stably by training in perceptual domain using a combination of multi-resolution STFT regression and adversarial losses.
\item We demonstrate the quality of the proposed model that outperforms LPCNet at a very low complexity of \SI{1.2}{GFLOPS} according to subjective and objective scores.

\end{itemize}

%% file: sections/fwgan.tex
Most GAN vocoders can be categorized into \textit{WaveNet-based} and \textit{latent-based} models. In WaveNet-based models, all network parameters are used at the same rate as the target signal. That is, if we seek to generate speech signal at sampling rate $f_{s}$, then every parameter in the model will do $f_s$ multiplications plus $f_s$ accumulations as the total number of floating-point operations per second (FLOPS). For light-weight vocoders such as Parallel WaveGAN~\cite{pwgan} with \SI{1.44}{M} parameters, generating one second of speech at \SI{16}{KHz} would require \SI{46}{GFLOPS}.

Latent-based models alleviate this computational burden by building the network as a stack of upsampling layers; where each layer operates at different resolution starting from the acoustic feature rate until reaching the target sampling rate of the signal. This can achieve moderate computational cost in ranges of \SI{10}{GFLOPS}~\cite{9632750} and \SI{8}{GFLOPS}~\cite{liu2021basis}, but with sophisticated architecture setups to maintain high quality. 

Framewise WaveGAN is a first step towards running GAN vocoders in the time domain at the acoustic feature rate, without having to use upsampling layers which are the main source of high complexity. This is achieved by making the model generate one frame at a time. Normally, in WaveNet-based and latent-based models, the feature representations starting from the first layer until the last one are organized as tensors of \textit{[Batch\textunderscore dim, Channel\textunderscore dim, Temporal\textunderscore dim]}; with \textit{Temporal\textunderscore dim} equal to the target signal resolution at the output layer. In Framewise WaveGAN, all feature representations are organized as \textit{[Batch\textunderscore dim, Sequence\textunderscore dim, Frame\textunderscore dim]}, where \textit{Sequence\textunderscore dim} is equal everywhere to the acoustic feature resolution that is commonly much smaller than the signal one; and \textit{Frame\textunderscore dim} holds the representation of the target frame that is being generated. The final waveform is obtained by simply \textit{flattening} the generated frames at the model output. This leads to significant computational saving even with models of large memory footprint. 
\begin{figure}[htb]
\begin{minipage}[b]{1.0\linewidth}
  \centering
  \centerline{\includegraphics[width=1.0\linewidth]{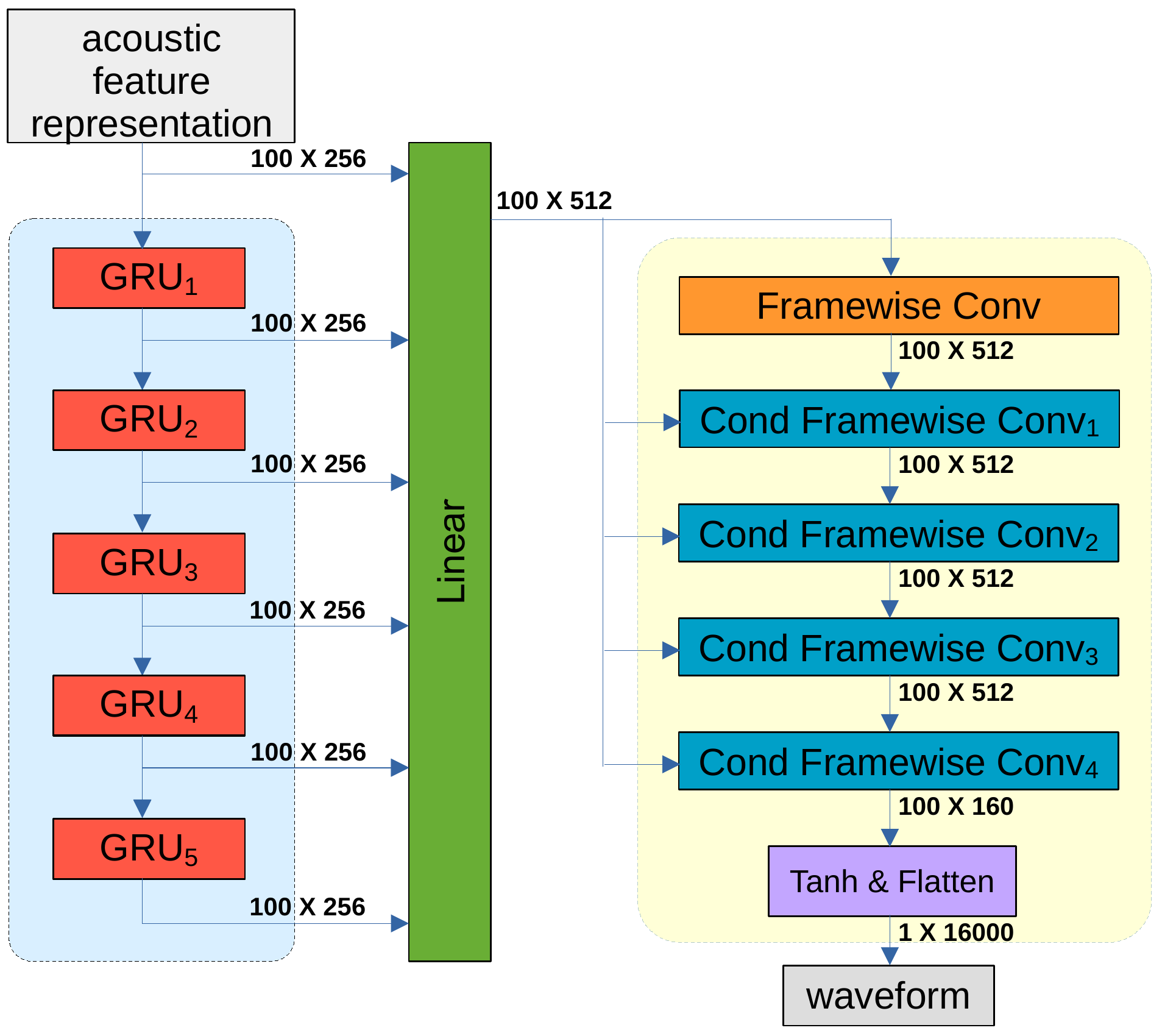}}
  \vspace*{0.3cm}
\end{minipage}
\caption{Framewise WaveGAN generator architecture. The numbers show \textit{[Sequence\textunderscore dim, Frame\textunderscore dim]} of the output representation from each layer to generate  \SI{1}{sec} of speech waveform at sampling rate of \SI{16}{kHz}, using conditioning acoustic features at \SI{100}{Hz}.}
\label{IMG_g}
\end{figure}
\subsection{Architecture Overview}
Fig.~\ref{IMG_g} illustrates the architecture of the proposed model. It mainly consists of two stacks of recurrent and fully-connected layers. The recurrent stack has 5~GRUs~\cite{chung2014empirical} to model long-term dependencies of the signal. All GRU outputs are concatenated with the conditioning (i.e., acoustic feature) representation and converted into lower-dimensional latent representation through a simple fully-connected layer. This representation is then utilized by the fully-connected stack that operate in framewise convolutional manner to model the short-term dependencies of the signal.  
\subsection{Framewise Convolution}
The term \textit{framewise convolution} refers to a kernel whose elements are frames instead of samples. We implement this by making the fully-connected layer receive at frame index $i$ a concatenation of $k$ frames at indices $\{i-k+1,..,i\}$ from the input tensor, where $k$ is the kernel size. The rest of the operation is same as normal convolution. There is also \textit{conditional framewise convolution} that only differs from framewise convolution in concatenating an external feature frame (i.e., conditioning vector) to the layer input. In our model, we use $1$ framewise convolution layer that receives the latent representation from the previous stack, with a kernel size of $3$ frames, stride $=$ dilation $=$ $1$ frame; and padding in non-causal manner (i.e., $1$ look-ahead frame). Hence, if the input tensor to this layer has \textit{Frame\textunderscore dim} of e.g., $512$, then the fully-connected network should have $3 * 512 = 1536$ input dimensions. In addition, there are $4$ conditional framewise convolution layers coming afterwards with a kernel size of $2$ frames which are concatenated with $1$ conditioning frame provided by the same latent representation obtained from the previous stack; with same stride, dilation and padding applied in causal manner. That's why the fully-connected network for this conditional layer has the same dimensionality as the non-conditional one. All of these framewise convolution operations are running in a single-channel sense; i.e., there is only one fully-connected network per layer. We used this way of implementation instead of traditional multi-channel convolution layers to ease the efficient implementation of the model, especially when applying sparsification methods to these layers. 
\subsection{Activation Layers}
For all layers in the recurrent and framewise convolution stacks, we use Gated Linear Unit (GLU)~\cite{dauphin2017language} to activate their feature representations:
\begin{equation}
GLU(X) = X \otimes \sigma(FC(X))\ ,
\end{equation}
where $FC$ is a simple fully-connected network to learn the sigmoid gate and it has the same output dimension as $X$, $\otimes$ denotes element-wise multiplication. We also disabled the bias for all layers in the model; which was helpful for faster convergence with lower reconstruction artifacts.
\subsection{Conditioning Acoustic Features}
We use LPCNet acoustic features~\cite{lpcnet} to condition our vocoder model. They consist of $18$ Bark-Frequency Cepstral Coefficients (BFCCs), a pitch period and a pitch correlation; which are extracted by \SI{20}{ms} overlapping windows at \SI{100}{Hz} frame rate from a \SI{16}{kHz} speech waveform. The model generates one \SI{10}{ms} frame per conditioning vector. The pitch period is fed to an embedding layer of $256$ levels and $128$ dimensions, while the BFCCs with the pitch correlation are fed to a simple causal convolution layer of $128$ output channels and kernel size of $3$. The outputs from these two layers are then concatenated and fed to another causal convolution layer of $256$ output channels, kernel size of $3$ and $Leaky\_ReLU$ activation to obtain the acoustic feature representation that is used for framewise generation, as shown in figure~\ref{IMG_g}. LPCNet features are calculated on \SI{10}{ms} frames with \SI{5}{ms} look-ahead; and Framewise WaveGAN requires one feature frame look-ahead. Hence, the total delay is \SI{10}{ms} for framing plus \SI{15}{ms} look-ahead, which sums to \SI{25}{ms}.

%% file: sections/training.tex
\subsection{Training in the Perceptual Domain}
Speech signals are characterized by their high dynamic range as they go wider in bandwidth. When we apply a simple pre-emphasis filter before training, we find the vocoder able to learn high frequency components faster than training in the normal signal domain. We reinforce this benefit by additionally using perceptual filtering, as detailed in AMR-WB~\cite{1175533}, so that the vocoder can learn high frequency content even faster. The perceptual weighting filter is defined by the following transfer function:
\begin{equation}
W(z) = \frac{A(z/\gamma_{1})}{(1 - \gamma_{2}z^{-1})},
\end{equation}
where $A(z)$ is the linear prediction (LPC) filter whose coefficients are computed from BFCCs, $\gamma_{1} = 0.92$ and $\gamma_{2} = 0.85$. This filtering increases the spectral flatness of signals during the training, which enables clearly faster convergence. Moreover, when applying inverse filtering to obtain the final signal, the noise of reconstruction artifacts is shaped by $W^{-1}(z)P^{-1}(z)$, where $P^{-1}(z)$ is the de-emphasis applied at end of the synthesis. The computational cost of this perceptual filtering is also quite cheap and still keeps the overall complexity low. 
\subsection{Spectral Pre-training}
We first pre-train the model using the spectral reconstruction loss $\mathcal{L}_\textup{aux}$ defined by Equation (6) in~\cite{pwgan}. It is a combination of spectral magnitude and convergence losses obtained by different STFT resolutions. We use all power-of-two FFT sizes between $64$ and $2048$ (6 sizes), with same values for window sizes and 75\% window overlap. For the spectral magnitude loss $\mathcal{L}_\textup{mag}$ defined by Equation (5) in ~\cite{pwgan}, we apply $sqrt$ instead of $log$ as a non-linearity, which was found better for early convergence. 
The spectral pre-training gives a metallic-sounding signal with over-smoothed high frequency content, which is a good prior signal to start adversarial training for achieving realistic signal reconstruction. 
\subsection{Spectral Adversarial Training}
Using time-domain discriminators is a major challenge in adversarial training of the proposed model. We tried different time-domain discriminator architectures~\cite{melgan,mustafa2021stylemelgan,kong2020hifi} that all failed to achieve stable training. Instead, we use multi-resolution spectrogram discriminators, which achieve much better training behavior and reliably increase the fidelity of generated signals. We follow the same spectrogram discriminator architecture defined in~\cite{jang21_interspeech} and use 6 models running on spectrograms of the same STFT resolutions used for spectral pre-training; with $sqrt$ used as a non-linearity.
The adversarial training uses least-square loss as a metric for evaluating discriminator outputs, with the same formulation given by Equations (1) and (2) in~\cite{kong2020hifi}. The spectral reconstruction loss is kept to regularize the adversarial training. Hence, the final generator objective is
\begin{equation}
\min_{G}\Big(\mathbb{E}_{z}\Big[\sum_{k=1}^6 (D_{k}\left(G(s)\right)-1)^{2}\Big]+\mathcal{L}_\textup{aux}(G)\Big),
\end{equation}
where $s$ represents the conditioning features (e.g., LPCNet features).
Weight normalization~\cite{salimans2016weight} is applied to all convolution layers of the discriminators ($D_{k}$) and all fully-connected layers of the generator ($G$).

%% file: sections/experiments.tex
\subsection{Experimental Setup}
We train the proposed Framewise WaveGAN model for a universal vocoding task, where the synthesis is speaker and language independent. To achieve this, we use a training speech corpus of $205$ hours sampled at \SI{16}{\kHz} and obtained from a combination of TTS datasets~\cite{demirsahin-etal-2020-open, kjartansson-etal-2020-open, guevara-rukoz-etal-2020-crowdsourcing,he-etal-2020-open,kjartansson-etal-tts-sltu2018,oo-etal-2020-burmese,van-niekerk-etal-2017,gutkin-et-al-yoruba2020,bakhturina2021hi}. The training data contains more than $900$ speakers in more than $34$ languages and dialects. Pre-emphasis with a factor of $0.85$ is applied to the speech signals when extracting the input LPCNet features. The complete synthesis is performed by running the vocoder model and then applying inverse perceptual filtering followed by de-emphasis. 
The training runs on an NVIDIA Tesla V100 GPU. We use batch size of $32$ and each sample in the batch corresponds to \SI{1}{sec} for both features and speech data tensors. The spectral pre-training is carried out for \SI{1}{M} steps (i.e., $\sim$ $50$ epochs), with $lr_g=10^{-4}$; and then the adversarial training runs for another \SI{1}{M} steps, with $lr_d= 2 * 10^{-4}$ and $lr_g$ is reduced to $5*10^{-5}$. AdamW~\cite{loshchilov2017decoupled} is used for both generator and discriminator optimizers, with $\beta = \{0.8, 0.99\}$. 
\subsection{Complexity}
In Framewise WaveGAN, all model parameters are used at the same rate of the conditioning features, which is \SI{100}{Hz}. Thus, each parameter contributes to one multiply and one accumulation (i.e., $2$ floating-point operations) per generation step. The total computational complexity to generate one second is then given as follows:
\begin{equation}
    C = N * 2 * S, 
\end{equation}
where $C$ denotes the number of floating-point operations per second (FLOPS), $N$ is the total count of model parameters and $S$ is the number of generation steps to create one second. Framewise WaveGAN according to the architecture defined in sec~\ref{fwgan} has \SI{7.8}{M} parameters. This gives a total complexity of \SI{1.5}{GFLOPS}, including \SI{7.3}{MFLOPS} for calculating $tanh$ and $sigmoid$ functions. We further reduce this complexity by sparsifying the dense model, as done in~\cite{wavernn, lpcnet}, with weight densities of $0.6$ for all GRUs and $0.65$ for all fully-connected layers except the last three, which are kept dense. This decreases the number of active parameters to \SI{5.9}{M} and makes the total complexity \SI{1.2}{GFLOPS}. We also checked the parallel generation speed for the Pytorch implementation of Framewise WaveGAN at \SI{1.2}{GFLOPS} in real-time factor (RTF). The model runs \SI{20}{x} and \SI{75}{x} faster than real-time on CPU (Intel Xeon Platinum 8175M 2.50GHz) and GPU (Nvidia Tesla V100), respectively.
\subsection{Quality}
We evaluate the model on the PTDB-TUG speech corpus~\cite{PirkerWPP11} and the NTT Multi-Lingual Speech Database for Telephonometry. From PTDB-TUG, we use all English speakers (10~male, 10~female) and randomly pick 200~concatenated pairs of sentences. For the NTT database, we select the American English and British English speakers (8~male, 8~female), which account for a total of 192~samples (12~samples per speaker). The training material did not include any data from the datasets used in testing. For comparison, we use the latest published low-complexity LPCNet~\cite{subramani2022end}. In addition, we also evaluate the reference speech as an upper bound on quality, and we include the Speex 4kb/s wideband vocoder~\cite{valin2007speex} as an anchor. We evaluate the models by the mean opinion score (MOS) results, obtained according to the crowd-sourcing methodology described in P.808~\cite{P.808}. To check how the quality scales with complexity, we test two LPCNet models at \{3, 1.2\} GFLOPS against three Framewise WaveGAN vocoders: the dense and sparse baseline models at \{1.5, 1.2\} GFLOPS; and a higher-complexity model of \SI{3}{GFLOPS}, whose architecture includes GRU size of $320$ and $10$ framewise convolution layers. 
\begin{figure}[htb]
\begin{minipage}[b]{1.0\linewidth}
  \centering
  \centerline{\includegraphics[width=1.0\linewidth]{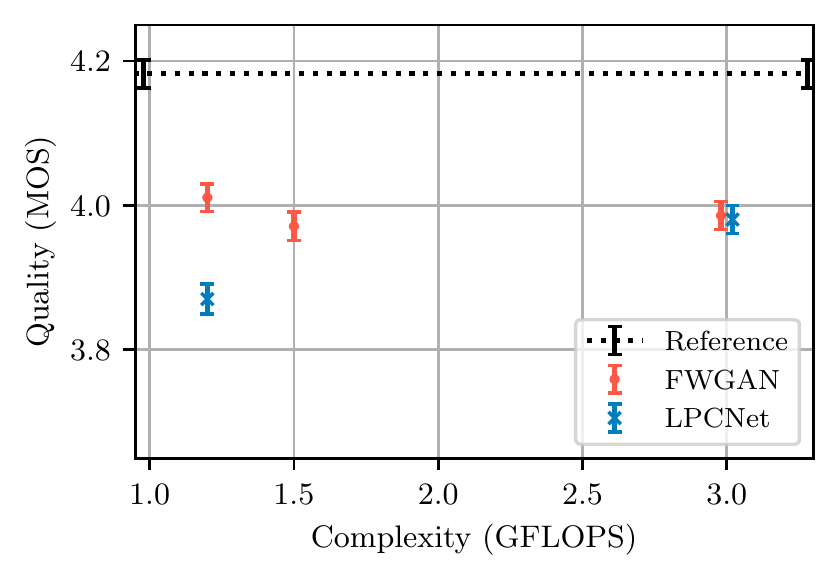}}
  \vspace{0.3cm}
\end{minipage}
\caption{Results from the MOS quality evaluation with 95\% confidence interval for LPCNet and Framewise WaveGAN (FWGAN) at different complexities.}
\label{mos}
\end{figure}

Fig.~\ref{mos} demonstrates the MOS test results performed by $30$ listeners using Amazon Mechanical Turk. Framewise WaveGAN (FWGAN) at \SI{1.2}{GFLOPS} achieves significantly higher quality than LPCNet at the same complexity\footnote{See our demo samples at the following url: \url{https://ahmed-fau.github.io/fwgan_demo/}}, even slightly outperforming (statistically significant) LPCNet operating at \SI{3}{GFLOPS}. On the other hand, the higher-complexity FWGANs cannot outperform the \SI{1.2}{GFLOPS} model. This is mainly due to the weaker discriminator behavior in adversarial training against generators with higher number of parameters. This may motivate further future work to come up with better discriminator models that enable quality scaling with higher generator complexities.
We also found Framewise WaveGAN has clearly better pitch consistency than the LPCNet model, as shown in Table~\ref{objective}; where PMAE is the pitch mean absolute error between ground truth and vocoder samples, VDE is the voicing decision error (i.e., the percentage of frames with incorrect voicing decision). We use YAAPT pitch tracker~\cite{5743729} for extracting the voicing information. Although we do not target singing voice generation in this work, this finding may encourage using Framewise WaveGAN for building low complexity expressive speech synthesizers.  
\begin{table}[th]
\caption{Objective evaluation of voicing features, lower is better.}
\label{objective}
\vspace{-4mm}
\begin{center}
\begin{tabular}{ l|c|c } 
\hline
\rule{0pt}{3ex} 
\textbf{Model} & \textbf{PMAE} & \textbf{VDE} \\
\hline
   
 LPCNet 3 GFLOPS &  $5.5865$ & $0.0168$\rule{0pt}{3ex}  \\ 
 LPCNet 1.2 GFLOPS &  $6.0965$ & $0.0177$ \\  
 FWGAN 1.2 GFLOPS  &  $5.0632$  & $0.0163$ \\ 
 FWGAN 1.5 GFLOPS  &  $5.3502$  & $0.0175$ \\ 
 FWGAN 3 GFLOPS  &  $5.4733$  & $0.0169$ \\ 
\hline
\end{tabular}
\end{center}
\end{table}